\documentclass[secnumarabic,amssymb, nobibnotes, aps, prl, preprint]{revtex4-1}
\usepackage[dvipdfm, dvips]{graphicx}% Include figure files
\usepackage{amsmath, mathtools}

\begin{document}

\title{Lattice deformation coupling of the electro-optic Kerr effect in liquid crystalline cholesteric blue phases}%
%observed by two-beam interference microscopy

\author{Hiroyuki Yoshida}
\altaffiliation[Also at: ]{PRESTO, Japan Science and Technology Corporation (JST), 4-1-8 Honcho Kawaguchi, Saitama 332-0012, Japan }
\email[e-mail: ]{yoshida@eei.eng.osaka-u.ac.jp}
\affiliation{Division of Electrical, Electronic and Information Engineering, Graduate School of Engineering, \\Osaka University, 2-1 Yamada-oka, Suita, Osaka 565-0871, Japan}

\author{Shuhei Yabu}
\author{Hiroki Tone}
\affiliation{Division of Electrical, Electronic and Information Engineering, Graduate School of Engineering, \\Osaka University, 2-1 Yamada-oka, Suita, Osaka 565-0871, Japan}

\author{Hirotsugu Kikuchi}
\affiliation{Institute for Materials Chemistry and Engineering, Kyushu University, Kasuga, Fukuoka 816-8580, Japan}

\author{Masanori Ozaki}
\affiliation{Division of Electrical, Electronic and Information Engineering, Graduate School of Engineering, \\Osaka University, 2-1 Yamada-oka, Suita, Osaka 565-0871, Japan}

\date{\today}
\begin{abstract}
The electro-optic Kerr effect in cholesteric blue phase liquid crystals is known to occur on sub-millisecond time scales, which is much faster than director reorientation in nematic liquid crystals.  Using two-beam interference microscopy, we report the presence of a very slow response in the Kerr effect, with a characteristic time of several seconds or more.  Using a simplified model for the reorientation dynamics, we attribute the slow response to the coupling between the local director reorientation and field-induced deformation of the lattice.  We provide evidence for our argument by showing that the slow response can be removed by inhibiting lattice deformation through polymer stabilization.
\end{abstract}
\maketitle

Liquid crystalline cholesteric blue phases (BPs) typically appear between the cholesteric phase and the isotropic liquid in a chiral liquid crystal (LC) \cite{wright_crystalline_1989,kitzerow_blue_2006}.  The cubic orientational order exhibited by BPs I and II makes them interesting both as subjects of soft matter physics \cite{fukuda_novel_2010,fukuda_quasi-two-dimensional_2011,fukuda_structural_2011,tiribocchi_switching_2011} and as candidate materials for next-generation electro-optic applications \cite{cao_lasing_2002,coles_liquid_2005,chen_reversible_2011,yabu_polarization-independent_2011}.  It is known that when a field is applied, both the refractive index and Bragg reflection wavelength change with a quadratic dependence on the electric field \cite{kitzerow_effect_1991,kikuchi_liquid_2008}, and using either of these properties, applications such as displays, tunable lenses, and reflectors have been proposed.  Furthermore, recent studies on BP--colloidal nanocomposites have opened the possibility of realizing a self-assembled, tunable metamaterial operative in the optical regime \cite{yoshida_nanoparticle-stabilized_2009, ravnik_three-dimensional_2011,yoshida_phase-dependence_2013}.

\begin{figure}[htb!]
 \centering
 \includegraphics[width=7cm]{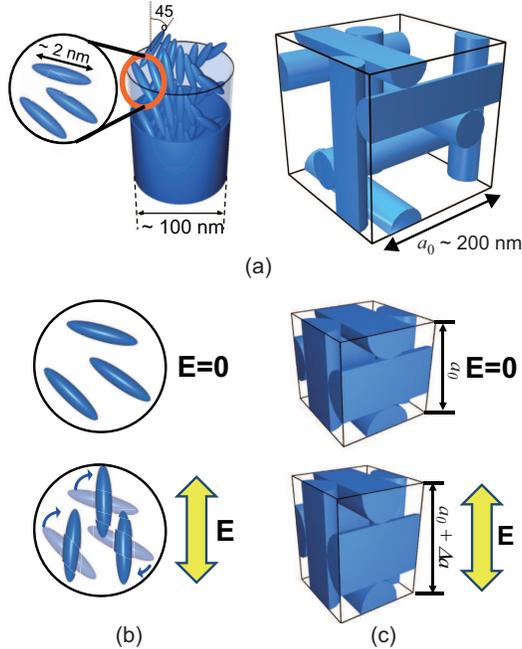}
 \caption{(a) Structure of body-centered BP I, which is characterized by a doubly twisted cylindrical structure, (b) local reorientation of shortly correlated nematic domains, giving rise to the electro-optic Kerr effect, and (c) electrostriction of the lattice.}
 \label{fig:modelBP}
\end{figure}

The two electro-optic effects observed in BPs, namely, the Kerr effect (which causes refractive index modulation) and electrostriction (which causes a change in the Bragg reflection wavelength), result from the same phenomenon, i.e., molecular reorientation, manifest at different temporal and spatial scales.  Figure \ref{fig:modelBP} illustrates the structure of BP I and the two electro-optic effects observed.  The Kerr effect is due to local reorientation of shortly correlated nematic domains existing in the BP lattice: because of the small helical pitch, the effect occurs at submillisecond response times, which is faster than the reorientation dynamics in nematics by about an order of magnitude \cite{gerber_electro-optical_1985, singh_measurement_1990,choi_electrooptic_2011, yabu_dual_2011}.  On the other hand, electrostriction of the lattice occurs so as to relax the increase in the elastic energy gained by local reorientation of molecules.  The effect involves as many as $10^7$ LC molecules, and consequently, has a characteristic time much longer than that of the Kerr effect, ca. several seconds or more.\cite{kitzerow_dynamics_1990} 

Considering the origin of these two electro-optic effects, it is intuitive to believe that they are interrelated with each other.  
However, to our knowledge, most studies to date have focused on examining the two effects independently and have used different cell configurations for each measurement.  Namely, the Kerr effect has been investigated by applying a horizontal electric field and measuring the induced birefringence \cite{gerber_electro-optical_1985, singh_measurement_1990, choi_electrooptic_2011}, and electrostriction has been investigated by applying a vertical field and measuring the shift in the Bragg reflection wavelength \cite{kitzerow_dynamics_1990}.  Gaining a unified understanding of the electro-optic response of BPs is important for potential future application of these materials.  Here, we use two-beam interference microscopy, which allows measurement of the transmitted phase and polarized reflection spectra to probe the dynamics of the two electro-optic effects simultaneously.  For the first time, to our knowledge, an extremely slow change in the refractive index is reported, which we attribute to the fact that the two effects are indeed coupled to each other.  We support our claim by the following two pieces of evidence: (i) a model incorporating a coupling coefficient between the two electro-optic effects can reproduce the observed experimental data, and (ii) the slow response can be removed by inhibiting lattice deformation though polymer stabilization.

The BP sample used in this study was prepared by adding a chiral dopant \lbrack ISO-(6OBA)$_2$, 7 wt\%\rbrack to a nematic LC mixture (5CB, 46.5 wt\% and JC-1041XX, 46.5 wt\%) with positive dielectric anisotropy \cite{shibayama_dendron-stabilized_2013}. We also prepared a polymer-stabilized (PS) BP by doping two types of monomers (dodecyl acrylate, 4.1 wt\% and RM257, 4.2 wt\%) and a photoinitiator (DMOAP, 0.8 wt\%) in the aforementioned BP sample \cite{kikuchi_polymer-stabilized_2002,haseba_large_2005}.  Sandwich cells with an approximate thickness of 27 $\mu$m, assembled from two pre-cleaned, indium tin oxide- (ITO-)coated glass substrates were filled with the samples.  The phase sequences of the samples were determined from polarized optical microscopy to be cholesteric (45.4 $^\circ$C)/BP I (45.9 $^\circ$C)/BP II (46.9 $^\circ$C)/isotropic for the BP (on heating the sample at a rate of 0.1 $^\circ$C/min) and cholesteric (30.0 $^\circ$C)/BP I (33.5 $^\circ$C)/BPII (35.5 $^\circ$C)/isotropic for the PSBP (on cooling the sample at a rate of 0.2 $^\circ$C/min).  The PSBP sample was polymerized at 32.5 $^\circ$C by irradiation with UV light (1.66 mW/cm$^2$, 365 nm) for 20 min, after which BP I was stabilized to below -60 $^\circ$C \cite{kikuchi_polymer-stabilized_2002}.  Electro-optic measurements were performed at 45.7 $^\circ$C for BP I and 27.5 $^\circ$C for PSBP.  

A two-beam interference microscope was built on a commercially available upright microscope (Olympus, BX-51).  An unpolarized He--Ne laser ($\lambda=632.8$ nm) was used for interferometry: it was divided by a non-polarizing cube beam splitter and then passed through either the sample path or the reference path.  In the sample path, the sample, an objective lens, and a half mirror were placed, whereas in the reference path, an achromatic doublet lens was inserted.  The two beams were recombined and imaged onto a CMOS camera equipped with a laser cleanup filter.  The sample was observed at the edge of the ITO electrodes so that regions both subject to and not subject to the field could be observed: the change in the refractive index was evaluated from the difference in the phase between the two regions.  The resolution of the setup, limited by fluctuations in the fringe originating from environmental vibrations, was $\delta n_{\rm{min}}\sim$ 10$^{-4}$.  The polarized reflection spectra were measured simultaneously using an incoherent white light source in the visible region ($\lambda$ = 400--800 nm).  A notch filter was inserted in the detection path to prevent the laser light from entering the spectrometer (Ocean Optics, USB-4000).  The diameter of the measured spot was $\sim$130 $\mu$m, and measurement was performed on the (110) platelet of BP I, which was confirmed by Kossel line observations.    The experiment was computer-controlled to acquire the interference fringe and the reflection spectrum at 200 ms intervals for 60 s.  A rectangular electric field (1 kHz) was applied for 30 sand then removed, and the voltage was incremented after a rest period of 60 s.  At this temperature, BP I showed a field-induced transition to the centered-tetragonal BP X at an applied fields above 2.7 V/$\mu$m \cite{yoshida_electro-optics_2013}: data were analyzed below this transition threshold.   

\begin{figure}[htbp]
 \centering
 \includegraphics[width=7cm]{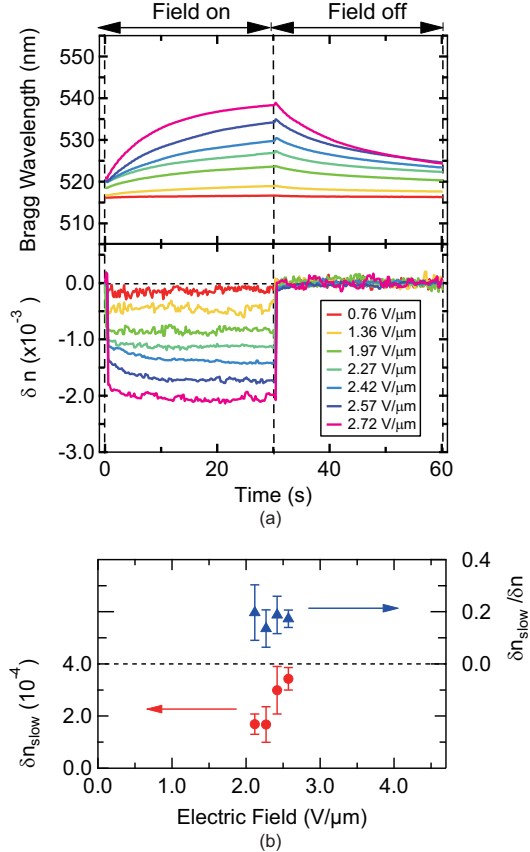}
 \caption{(a) Transient response of Bragg wavelength and change in refractive index for BP I sample, and (b) applied field dependence of slow change in refractive index.}
 \label{fig:resultBPI}
\end{figure}

Figure \ref{fig:resultBPI} shows the transient response of the Bragg wavelength and relative refractive index of the BP sample.  The Bragg wavelength shows a relatively slow response with characteristic times of 10 s or more: this is because the shift in the Bragg wavelength is due to an elastic deformation of the lattice involving a large number of LC molecules \cite{kitzerow_dynamics_1990}.  The refractive index, on the other hand, shows a faster response because it is caused mainly by reorientation of the nematic director on a scale smaller than the lattice constant \cite{gerber_electro-optical_1985}.  As reported previously in many studies, the response time is generally on the submillisecond scale and is observed as a step-like response near 0 and 30 s in the time resolution of this measurement \cite{singh_measurement_1990,choi_electrooptic_2011,yabu_dual_2011}.  Surprisingly, however, for electric field intensities between 2.1 and 2.6 V/$\mu$m, a very slow change in the refractive index is observed, with characteristic times similar to that of electrostriction.   The presence of this response was confirmed by making at least five measurements, and was found to be on the order of 10$^{-4}$, which was as large as 20 \% of the total shift in the refractive index.  In Fig. \ref{fig:resultBPI}(b), we show the applied field dependence of the slow change in the refractive index $\delta n_{slow}$: although the effect was not observable at low fields because of the limitation of our setup, a monotonic increase was observed between applied fields of 2.1 and 2.6 V/$\mu$m.  

We believe that the unusually slow change in the refractive index is actually a natural consequence of the fact that the two electro-optic effects originate from the same phenomenon of molecular reorientation and are therefore coupled.  This coupling phenomenon can be described physically as follows.  The Kerr coefficient has been shown to have a cubic dependence on the BP lattice constant \cite{choi_electrooptic_2011}.  Thus, the extent to which the local nematic director can reorient is limited by the three-dimensional structure of the BP.   As the lattice deforms under an applied field, the molecules are able to reorient further in the direction of the field, which in turn deforms the lattice further, until equilibrium is reached.  The extra reorientation of the LC molecules enabled by lattice deformation contributes to an extra change in the refractive index with a characteristic response time similar to that of electrostriction.  Interestingly, the slow response is observed only when the field is applied to the sample and not when the field is removed.  This does not contradict the description above, since there is no directional torque acting in the off-switching process.  This indicates that the molecules can return instantaneously to their initial helical configuration, such that the difference in the refractive index is negligible.  

We present a model that describes the dynamic response of the two electro-optic effects.  To fully describe the reorientation dynamics of BPs, simulation of the order parameter tensor would be necessary; to prove our point, we employ here a coarse-grained approach in which the free energy density is described in terms of the effective order parameter $S_{{\rm BP}}$ and electrostriction $u$ of a single unit cell of the BP lattice, i.e.
\begin{eqnarray}
\label{eq:def}
\begin{split}
S_{{\rm{BP}}}  &=  \frac{1}{{V_{{\rm{BP}}} }}\int_{V_{{\rm{BP}}} } {\frac{1}{2}\left( {3\cos ^2 \theta  - 1} \right)dV}\\
u &= \frac{{d\left( E \right) - {d_0}}}{{{d_0}}}.
\end{split}
\end{eqnarray}
$S_{\rm {BP}}$ is related to the induced birefringence through the relationship $\Delta n_\textrm{ind} = - \delta n/3  = {S_{{\textrm{BP}}}} \times \Delta n_\textrm{LC}$, where $\Delta {n_{{\rm{LC}}}}$ is the intrinsic birefringence of the LC molecules, which is determined to be 0.076 by measuring the phase after applying an electric field with an intensity twice the BP--nematic transition threshold ($E_{\rm {critical}}=7.8$ V/$\mu$m).  In Eq. \ref{eq:def}, $\theta$ is the angle between the local director and the electric field, $V_{\rm{BP}}$ is the volume of the unit cell, $d\left( E \right)$ is the periodicity along the $\langle$110$\rangle$ direction at an applied field of $E$, and $d_0$ is the periodicity without the field, which is related to the cubic lattice constant $a_0$ through the relation $d _0 =\sqrt{2}a_0$.   Because of the cubic symmetry of the BP, $S_{{\textrm{BP}}} = 0$ and $u = 0$ at zero applied field: upon application of a field, both values increase monotonically with the field intensity.   Assuming that $S_{{\rm BP}}$ and $u$ experience an elastic restoring force with proportionality constants $\kappa_1$ and $\kappa_2$, respectively, we may write the free energy density as follows:
\begin{equation}
f = \frac{1}{2}\kappa _1 \left( {S_{{\rm{BP}}}  - cu} \right)^2  + \frac{1}{2}\kappa _2 u^2  - \frac{1}{2}\varepsilon _0 \Delta \varepsilon S_{{\rm{BP}}} E^2.
\end{equation}
The first and second terms describe the gain in free energy caused by local molecular reorientation and electrostriction, respectively, and the third term is the contribution of the electric energy.  The parameter $c$ in the first term is a coupling coefficient we introduce to account for the effect of lattice deformation on the local director reorientation, and is a measure of the additional director reorientation enabled by lattice deformation.  

The Euler-Lagrange equations for $S_{\rm{BP}}$ and $u$ yield the equations of motion for the system, where $\eta _1$ and $\eta _2$ are parameters related to the viscosities of each effect.  
\begin{gather}
\label{eq:euler-lagrange}
\begin{split}
 \eta _1 \frac{{dS_{{\rm{BP}}} }}{{dt}} =  - \kappa _1 \left( {S_{{\rm{BP}}}  - cu} \right) + \frac{1}{2}\varepsilon _0 \Delta \varepsilon E^2,  \\ 
 \eta _2 \frac{{du}}{{dt}} = c\kappa _1 \left( {S_{{\rm{BP}}}  - cu} \right) + \kappa _2 u.
 \end{split}
 \end{gather}
The solution to the above simultaneous differential equation is a double-exponential function with time constants that are determined primarily by $\tau _1 = \eta _1 / \kappa _1$ and $\tau _2 = \eta _2 / \kappa _2$, but modulated by the presence of the coupling coefficient $c$.  Considering that the dominant response observed by interference microscopy is electrostriction, we may describe the experimentally obtained response curves using only a single exponential function,
\begin{gather}
\label{eq:fit_func}
\begin{split}
\delta {n_{{\rm{slow}}}}\left( t \right) = {A_{\delta n}}\exp \left( { - t/{\tau _u}} \right) + C_{\delta n},\\
u\left( t \right) = {A_u}\exp \left( { - t/{\tau _u}} \right) + {C_u},
\end{split}
\end{gather}
where $A_{\delta n, u}$ and ${C_{\delta n, u}}$ are constants, and $\tau _u$ is the time constant of the slower response.  By comparing the steady-state solutions from Eq. \ref{eq:euler-lagrange} with Eq. \ref{eq:fit_func}, the coupling coefficient $c$ can be determined experimentally by the following expression:
\begin{equation}
\label{eq:coef}
c = \frac{{3{A_{\delta n}}}}{{\Delta {n_{{\rm{LC}}}}{C_u}}}.
\end{equation}

\begin{figure}[htbp]
 \centering
 \includegraphics[width=7cm]{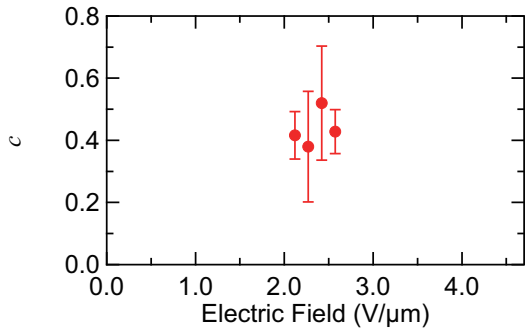}
 \caption{Coupling coefficient obtained for various applied electric field intensities.  }
 \label{fig:BP_C}
\end{figure}

Figure \ref{fig:BP_C} shows the values of $c$ obtained when various electric fields were applied to the sample.  The experimental results are reproduced when values of approximately 0.4 are assumed for the coupling constant.  Furthermore, we performed the same experiments on a BPLC prepared from a different host and chiral dopant (Merck, E44, 40 wt\% and CB15, 60 wt\%), and obtained values close to 0.2.  We therefore believe that the coupling of the Kerr effect with electrostriction is a universal feature in BPs, and that the coupling magnitude depends on the physical parameters of the LC.  

Because director reorientation results from competition between the elastic energy of the LC and the dielectric energy of the applied field, it is likely that the coupling coefficient depends on the elastic constant of the liquid crystalline molecules.  Moreover, it was shown recently that the anisotropy of the elastic constant, or more specifically, the ratio of the bend elastic constant to the splay elastic constant $K_{33}/K_{11}$, has a profound impact on the stability and electrostriction of BPs \cite{fukuda_stabilization_2012,shibayama_dendron-stabilized_2013}.   The difference in the coupling coefficients of the two materials investigated may therefore be a consequence of  different anisotropies in their elastic constants.  However, because of our simplified approach in modeling the phenomenon, the coupling coefficient cannot be associated directly with the physical parameters of the LC.  Geometrical factors such as the cell-gap and surface anchoring are also neglected in our discussion.  Although the discussion above allows us to show that the slow response is a consequence of the coupling between the Kerr effect and electrostriction, it would be illuminating to perform numerical simulations in the future on the dynamics of the order parameter tensor incorporating anisotropic elastic constants.

\begin{figure}[htbp]
 \centering
 \includegraphics[width=7cm]{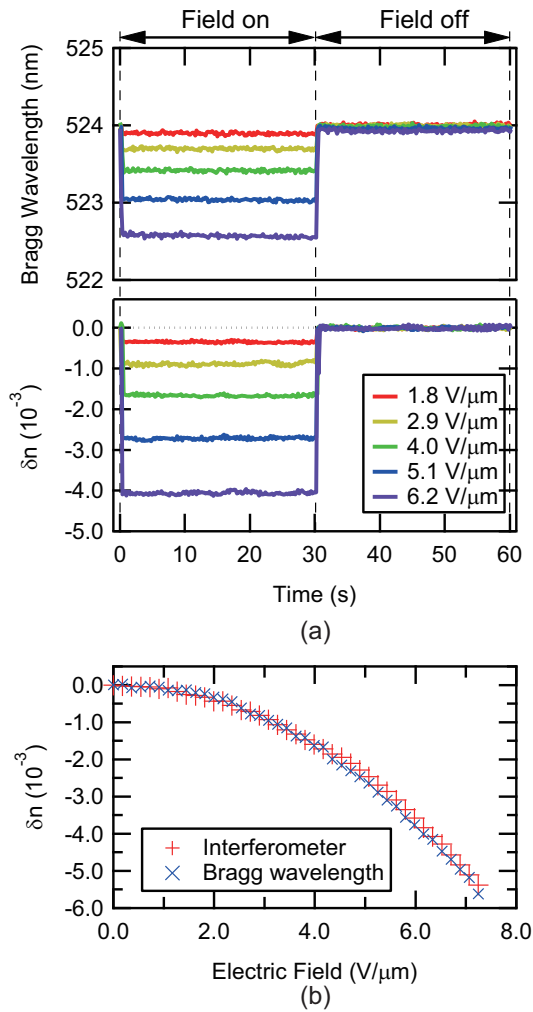}
 \caption{(a) Transient response of  Bragg wavelength and change in refractive index for PSBP, and (b) change in refractive index measured by interferometry and spectroscopy.}
 \label{fig:PSBP}
\end{figure}

We provide further evidence that the slow change in the refractive index is caused by coupling of the local director reorientation to the deformation of the lattice by showing that the slow response can be removed completely by stabilizing the lattice structure.  PSBPs are a class of BP materials in which the lattice structure is frozen by a fine network of polymer templating the periodic structure but that can still show a Kerr-type electro-optic response because of local molecular reorientation of the molecules existing among the polymer network \cite{kikuchi_polymer-stabilized_2002, haseba_large_2005}.  Therefore, if the slow response of the refractive index originates in lattice deformation, it should not be observed in the PSBP.  Figure \ref{fig:PSBP}(a) shows the transient response of the PSBP at various field intensities.  As expected, the slow responses of both the Bragg reflection wavelength and refractive index vanish, thereby providing direct evidence that lattice deformation coupling of the electro-optic Kerr effect is responsible for the slow response.  

Interestingly, in contrast to the behavior of the non-polymerized BP sample, the Bragg wavelength shows a small blue shift upon application of a field.  This can be attributed to the change in the refractive index of the sample, since Bragg reflection from the (110) plane of a cubic lattice occurs at $\lambda = n a_0/\sqrt{2}$.  A blue shift of the reflected wavelength implies that the refractive index has decreased under the applied field, which is consistent with the refractive index modulation induced by the Kerr effect.  Figure \ref{fig:PSBP}(b) compares the electric field dependence of the relative refractive index obtained from interferometry and spectroscopy: the two values match almost perfectly, indicating that the small peak shift is indeed due to the Kerr effect.  

To conclude, we employed two-beam interference microscopy to probe the electro-optic response in BP and PSBP LCs.  The BP showed a slow change in refractive index with a characteristic response time comparable to that of electrostriction, whereas no such response was observed in the PSBP with a fixed lattice structure.  The slow response occurs because the electro-optic Kerr effect is coupled to electrostriction, which originates from the fact that the two electro-optic effects arise from the same phenomenon of molecular reorientation.  That the very structure that gives rise to the attractive features of BPs is the cause of the slow response poses a new challenge for the practical application of BPs.  Recent studies have focused on developing material systems with wide BP stability ranges \cite{coles_liquid_2005, yoshizawa_binaphthyl_2009, yoshida_nanoparticle-stabilized_2009, karatairi_nanoparticle-induced_2010, kogawa_biphenyl_2011, shibayama_dendron-stabilized_2013}, and the dynamics of the electro-optic response is often overlooked.  We hope that this study will raise awareness of this subject and promote experimental and theoretical studies targeted at clarifying the mechanisms further and minimizing the effect.   

\acknowledgements{This work was supported by JSPS KAKENHI Grant Numbers 24656015, 23656221, and 23107519. H. Yoshida acknowledges financial support from the JST PRESTO Program.}

\bibliography{bibfile}

\end{document}